\documentclass[thmsa,a4paper]{article}
\usepackage{amssymb}
\usepackage{sw20asas}

\input tcilatex
\begin{document}

\begin{center}
* * *

A CRUCIAL DIPOLE TEST OF THE EXPANSION CENTER UNIVERSE

BASED ON HIGH-$Z$ SCP UNION \& UNION2 SUPERNOVAE

\vspace{.3in}

ECM paper XV by Luciano Lorenzi

by merging the SAIt 2011 ECM paper X with the EWASS 2012 ECM paper XII

\vspace{.3in}

ABSTRACT
\end{center}

\begin{quotation}
\textbf{The expansion center Universe (ECU) gives a dipole anisotropy to the
Hubble ratio, as }$cz/D=H_0-a^{*}\cos \gamma $ ,\textbf{\ at any Hubble
depth }$D$\textbf{. After a long series of successful dipole tests, here is
a crucial multiple dipole test at }$z$\textbf{\ bins centred on the mean }$%
\langle z\rangle \equiv z_0=1.0$,\textbf{\ or Hubble depth }$D=c/H_0$\textbf{%
, and based on data from SCP\ Union \& Union2 compilation. Table 5abc lists
data of two main samples, with 48 }$SCPU$\textbf{\ SNe Ia and 58 }$SCPU2$%
\textbf{\ SNe Ia respectively. The confirmed dipole anisotropy, shown by }$6$%
\textbf{\ primary sample tests and by another }$27$\textbf{\ from }$9$ 
\textbf{encapsulated }$z$ \textbf{bins with }$D_L=D\cdot (1+z)$\textbf{\
assumed and the Hubble Magnitude definition},\textbf{\ gives the mean }$%
\langle a^{*}\rangle \cong 5.5\pm 0.3$ $km/s/Mpc$ \textbf{as a model
independent result, in full accordance with the expansion center model
(ECM). That means a maximum }$cz$ \textbf{range of }$\sim 50000$ $km/s$ 
\textbf{at }$z_0=1$\textbf{, that is }$c\left| \Delta z\right| ^{\max
}=2a^{*}D$\textbf{\ with a decelerating expansion dipole coefficient }$%
a_{ECM}^{*}\cong 5.46$\textbf{\ H.u. at }$D\cong 4283$ $Mpc$\textbf{. As a
complement to the dipole tests, here is a new computation of the
relativistic deceleration parameter }$q_0$\textbf{, based on the
extrapolated total }$M$\textbf{\ spread, that is the deviation of the Hubble
Magnitude }$M$\textbf{\ of high-}$z$\textbf{\ SCP Union supernovae at a
normal or central redshift }$\langle z\rangle \equiv z_0=z$ $\ll 1$ \textbf{%
from the absolute magnitude }$M_0$ \textbf{at }$z_0\rightarrow 0$\textbf{\
(cf. parallel paper XVI). A total }$M$\textbf{\ spread according to ECM is
derived from 249 high-}$z$\textbf{\ }$SCPU$\textbf{\ SNe listed in paper
XVI. In a concordance test with the expansion center model, the obtained new
relativistic }$q_0\gtrsim +2$ \textbf{agrees with the value }$q_0=+2$\textbf{%
\ inferred from the ECM paper I eq. (41),} \textbf{when }$R_0$ \textbf{is
the proper distance at }$t_0$\textbf{\ of the expansion center from the
Galaxy.}
\end{quotation}

\newpage

\section{Introduction}

After the construction of the wedge-shaped Hubble diagram of 398 SCP
supernovae (Lorenzi 2010), the priority is to check the ECM dipole
anisotropy of the Hubble ratio $cz/D.$

Here a first crucial multiple dipole test has been carried out on the most
remote supernovae SNe Ia (a lot of which were observed with the NASA/ESA 
\textit{Hubble Space Telescope),} ranging around the mean value $\langle
z\rangle =1.0$ and listed in the SCP Union ($SCPU$ hereafter) compilation by
Kowalski et al. (2008) and in the SCP Union2 ($SCPU2$ hereafter) by
Amanullah et al. (2010).

Such a multiple dipole test is intended to check the expansion center
Universe (ECU) described by the fundamental equation (59) of paper I or eq.
(1) of paper II, without using the expansion center model (ECM) developed in
the ECM paper series and based on the Galaxy radial deceleration coefficient 
$a_0=K_0R_0$ of eq. (3) and (9) in paper II, with $K_0=\left( \frac{\delta H%
}{\delta r}\right) _0=\frac{3H_0^2}c$ as resulting in section 4 of paper I
and experimentally confirmed in the concluding section of the same paper I.
In other words the present contribution confirms the Hubble ratio dipole as
a model independent result, with a resulting angular coefficient which has
the same value as predicted by the ECM.

The present dipole test includes also the contents of the ECM paper XII,
''Evidence for a high deceleration of the relativistic Universe'', presented
at the European Week of Astronomy and Space Science (EWASS 2012). So paper
XV is a combination of paper X with paper XII.

Let us remark that the convention $M_B\equiv M$ is adopted in this paper XV,
while the cited papers I-II-III-IV-V-VI-VII-VIII-IX-X-XI-XII-XIII-XVI are
those referenced as Lorenzi 1999a$\rightarrow $2012e.

\section{The Hubble depth $D$ from the new Hubble law}

The first works on the expansion center Universe (ECU: Lorenzi 1989-91-93)
dealt with the trigonometric distance $r$ as the classic separation between
our Milky Way and the nearby galaxies / groups / clusters / superclusters,
at depths $z\lesssim 0.1$, with luminosity distances $D_L\equiv r$ assumed.
Such an approximation was more suitable in the ECM papers I-II (1999ab),
where the \textbf{light-space} $r$ of galaxies / groups / clusters with $%
z\lesssim 0.03$ was fixed mathematically as 
\begin{equation}
r=-c(t-t_0)
\end{equation}
that is the distance covered by light at a constant speed $c$ during the
whole travel time, from the emission epoch $t_e=t$ to the present epoch $t_0$%
. Since $c$ is constant, in eq. (l) $r$ should correspond to the source
distance at the emission epoch. However the ''cosmic medium'' (CM), with
respect to which light moves at speed $c$, is expanding as does the whole
Universe; as a result, the light-space $r$ is larger than the distance at
the emission epoch, although its value in light-time represents a measure of
that past epoch $t$. Now, while $r$ is unknown, its derivative with respect
to the emission time $t$, after putting $dt=\lambda /c$ and $dt_0=\lambda
_0/c$ with $\Delta \lambda =\lambda _0-\lambda $, becomes the observed and
well-known $cz$ (cf. papers V, VIII and IX), as follows: 
\begin{equation}
\dot r=\frac{dr}{dt}=c\frac{d(t_0-t)}{dt}=c\frac{\Delta \lambda }\lambda =cz
\end{equation}

Such $\dot r=cz$ enters the ECU \textbf{new Hubble law} (cf. papers I$%
\rightarrow $IX) in Hubble units: 
\begin{equation}
\dot r=Hr+\Delta H\cdot (r-R\cos \gamma )+R\dot w\sin \gamma
\end{equation}

Specifically $\gamma $ is the angle between the direction of the huge void
center $VC$ ($\alpha _{VC}\approx 9^h;$ $\delta _{VC}\approx +30^0:$ Bahcall
\& Soneira 1982), also called expansion center or Big Bang central point
(Lorenzi 1989-91-93) distant $R_0$ from the Local Group (LG) at our epoch,
and that $(\alpha ,\delta )$ of the observed outer
galaxy/group/cluster/supernova at a distance $r$ from LG, with the nearby
Universe radial velocity $\dot r$ corrected only by the standard vector
(Sandage \& Tammann 1975a) (Lorenzi 1999a: paper I).

Eq. (3), by leaving out the formulae of the related expansion center model
(cf. section 4 in paper I and section 2 in paper II) and assuming $\langle
\dot w\rangle =0$ (cf. section 7.4 in paper I), can be expressed by an
alternative formulation, which easily leads to the Hubble depth $D$. In this
case, putting $H_{*}=H+\Delta H$, eq. (3) may be rewritten in terms of the
following sequence: 
\begin{eqnarray}
cz &=&rH_{*}-R\Delta H\cos \gamma =DH_0-R\Delta H\cos \gamma \Rightarrow
r(H+\Delta H)\equiv DH_0\Rightarrow \\
\Delta H &=&\Delta H(D)\Rightarrow R\Delta H\equiv a^{*}(D)\cdot
D\Rightarrow cz=D\cdot H_0-a^{*}(D)\cdot D\cdot \cos \gamma
\end{eqnarray}
\begin{equation}
rH_{*}\equiv DH_0
\end{equation}

The identity (6) of the two products shows that here we have two physical
quantities, $r$ and $H_0$, and two apparent quantities, $H_{*}$ and $D$.
Consequently, as $H_{*\text{ }}$is the apparent Hubble constant of the
observed source at the emission epoch $t$, similarly $D$ \textbf{results to
be the apparent distance} of the observed source at the present epoch $t_0$,
that is, the \textbf{Hubble depth }$D$\textbf{\ of the ECU Hubble law,
according to the formulae reported below}. As $H_{*}>H$, thus $D>D_0$, where 
$D_0$ represents the proper distance of the observed source at the present
epoch $t_0$, while $H$ represents the physical Hubble constant of the
observed source at the emission epoch $t$. Based on the canonical
terminology, one should distinguish the use of ''proper'' and ''physical''
as quantity adjectives. Indeed the problem is with light-space, which, when
considered as the proper distance $r_{pr}$ referring to the emission epoch $%
t $, is shorter than the physical distance $r$ run by light during the whole
travel time. This is the reason for giving the adjective ''physical'' to the
light-space $r$, according to the identity (6).

Now, coming back to eqs. (3)(4)(5), here is the formulation for the
wedge-shape of the ECU Hubble law, or \textbf{new Hubble D law (7), }with a
few specifications: 
\begin{equation}
cz=\left[ H_0-a^{*}(D)\cos \gamma \right] \cdot D\Rightarrow c\left| \Delta
z\right| ^{\max }=2a^{*}D
\end{equation}
\begin{equation}
H_X=H_0-a^{*}(D)\cos \gamma
\end{equation}
\begin{equation}
a^{*}\neq 0\hspace{.2in}and\hspace{.2in}\cos \gamma =0\Rightarrow
cz=cz_0=H_0D
\end{equation}
\begin{equation}
D=cz/H_X=cz_0/H_0
\end{equation}
\begin{equation}
a^{*}\equiv 0\Rightarrow H_X=H_0
\end{equation}

Hence the new depth law clearly shows the anisotropic behaviour of $H_X$,
whose variability (owing to $a^{*}\cos \gamma $) is responsible for the
wedge shape of the Hubble diagram. Furthermore eq. (10) allows us to
represent the Hubble depth $D$ according to the powerful $h$ convention, as
follows: 
\begin{equation}
D=dh^{-1}Mpc\hspace{.4in}being\hspace{.2in}d=\frac{cz}{100\text{ }km\text{ }%
s^{-1}Mpc^{-1}}\hspace{.2in}and\hspace{.2in}h=\frac{H_X}{100\text{ }km\text{ 
}s^{-1}Mpc^{-1}}
\end{equation}

In particular at $z=1$ it results that $D=2998$ $h^{-1}Mpc$, while our
dipole test at $\langle z\rangle =1.0$, assuming $\langle z\rangle \equiv
z_0=z$ at $\cos \gamma =0$, refers to $z$ bins with the Hubble depth $%
D=D(z_0)$, that is 
\begin{equation}
H_0\equiv 70\text{ }km\text{ }s^{-1}Mpc^{-1}\text{ }and\text{ }\langle
z\rangle \equiv z_0=1.000\Rightarrow \cos \gamma =0\Rightarrow D=\frac
c{H_0}=4283\text{ }Mpc
\end{equation}

\section{Two dipole tests on SCP\ supernovae at $\langle z\rangle =1.0$}

After a long series of successful dipole tests on the nearby Universe
(Lorenzi 1991-93-94-99ab-2003b), from historic data sets of about half a
century, and that carried out on 53 SCP SNe Ia ranging around $\langle
z\rangle =$\ $0.5$\ (ECM paper VI based on data by Perlmutter at al. 1999
and Knop et al. 2003), here a crucial multiple dipole test at $z$\ bins with 
$\langle z\rangle \equiv z_0=1.0$\ or Hubble depth $D(z_0)=c/H_0\equiv 4283$ 
$Mpc$ is carried out on SNe Ia data and references from SCP Union
compilation ($SCPU:$\ Kowalski et al. 2008) and SCP Union2 ($SCPU2:$\
Amanullah et al. 2010), including the results obtained within ''The new
wedge-shaped Hubble diagram of 398 SCP supernovae according to the expansion
center model'' (ECM paper IX: SAIt2010 in Naples).

Assuming directly $H_0\equiv 70$ Hubble units, as the conventional ECM
Hubble constant derived from the value $H_0=69.8\pm 2.8$ H.u. in 1999 paper
II and based on data from Sandage \& Tammann (1975), means limiting the
fitting of the ECM dipole formula (8) of paper VI to one unknown, through
the first eq. (7) rewritten in the form 
\begin{equation}
Y=a^{*}(D)\cdot (-\cos \gamma )\hspace{.5in}with\hspace{.5in}Y=\frac{cz}D-H_0
\end{equation}
where the supernova Hubble depth $D$ is computed as the ratio between the
cosmological distance $D_C$ and $1+z$ , \textbf{with} $D_C\equiv D_L$ 
\textbf{assumed} (cf. papers V-VI-IX), that is by taking the position 
\begin{equation}
D_C=D\cdot \left( 1+z\right) \equiv D_L
\end{equation}
and a consequent magnitude formula that we call the \textbf{Hubble Magnitude}
(cf. section 3.1 and section 5), while the SNe angular position with respect
to the huge void center $VC$ is expressed by the usual $\cos \gamma $
formula: 
\begin{equation}
\cos \gamma =\sin \delta _{VC}\sin \delta +\cos \delta _{VC}\cos \delta \cos
(\alpha -\alpha _{VC})
\end{equation}

Table 0 presents the 4 pilot samples from which the useful samples at $%
\langle z\rangle =1.0$ were extracted. These pilot samples exclude $z\leq
0.2 $ in order to eliminate the CMB reference (cf. section 3 of paper IX).
In particular XVI refers to 249 SNe Ia selected by the SCP Union, those
listed in the Appendix of paper IX and lying within the Hubble depth range $%
800$ $Mpc<D<8000$ $Mpc$. XVII refers to all the 283 SNe Ia at $z>0.2$ of the
SCP Union. XVIII refers to 338 SNe Ia selected by the SCP Union2, also at $%
z>0.2$. XIX refers to all the 359 SNe Ia at $z>0.2$ of the SCP Union2. All
the selected data ranging around $\langle z\rangle =1.0$ , both from the SCP
Union and SCP\ Union2, have been listed in Table 5abc of section 3.4, as
they refer to two main samples, the XVI$_1$ with 48 $SCPU$\ SNe Ia and the
XVIII$_1$ with 58 $SCPU2$\ SNe Ia, respectively.

\vspace{.1in}

\textbf{Table 0}

\vspace{.1in}

\begin{tabular}{|c|c|c|c|}
\hline
Pilot sample & source & N & $z$ bin \\ \hline
XVI & Kowalski et al. 2008 $(SCPU)$ & 249 & $0.200<z\leq 1.390$ \\ \hline
XVII & Kowalski et al. 2008 $(SCPU)$ & 283 & $0.200<z\leq 1.551$ \\ \hline
XVIII & Amanullah et al. 2010 $\left( SCPU2\right) $ & 338 & $0.200<z\leq
1.400$ \\ \hline
XIX & Amanullah et al. 2010 $(SCPU2)$ & 359 & $0.200<z\leq 1.400$ \\ \hline
\end{tabular}

\vspace{.4in}

\subsection{1$^{st}$dipole test: $M=d_0+d_1D+d_2D^2$}

The first-type dipole test refers to the pilot samples XVI and XVII of Table
0. Indeed the construction of the new wedge-shaped Hubble diagram, in paper
IX, was based on the rigorous coincidence of the SNe Ia Hubble magnitude, as 
$M=m-5\log D_C-25$, with its computed average trend, $\langle M\rangle
=d_2D^2+d_1D+d_0$, according to the adopted relation (24) in paper IX, that
is 
\begin{equation}
M=m-5\log \left[ D\cdot (1+z)\right] -25\equiv d_2D^2+d_1D+d_0=\langle
M\rangle
\end{equation}

\textbf{We define Hubble Magnitude as the quantity }$(m-5\log \left[ D\cdot
(1+z)\right] -25)$ \textbf{in H.u., where }$D$ \textbf{is the Hubble depth
representing the apparent distance at the present epoch }$t_0$\textbf{\ in
the Hubble diagram, according to the position (15) and eq. (12).}

The introduction of the values $d_0,d_1,d_2,z,m$ in eq. (17) (assuming $%
d_0=-18.77;$ $d_1=-1.421\cdot H_0/c;$ $d_2=+0.3589\cdot H_0^2/c^2$ from
paper IX), gave the Hubble depth $D$ through a numerical solution point by
point. Hence, after calculating each $\cos \gamma $ according to eq. (16),
we obtain $Y$ and a system of equations (14), whose solution by means of the
least square method gives the unknown angular coefficient $a^{*}$. In
synthesis the first dipole test may be summarized as follows: 
\begin{equation}
\left[ d_0,d_1,d_2,z,m\right] \Rightarrow D\hspace{.5in}\text{ }\left[
z,D\right] \Rightarrow Y\hspace{.5in}\cos \gamma \Rightarrow \left[
Y\rightarrow (-\cos \gamma )\right] \Rightarrow a^{*}
\end{equation}

The above procedure is then applied to $10$ $z$ bins of the pilot sample XVI
(XVI$_{1\rightarrow 10}$), those having $\langle z\rangle \equiv z_0=1.0$,
and to the first $z$ bin of the pilot sample XVII (XVII$_1$)(cf. Tables
1ab-2).

\subsection{2$^{nd}$dipole test: $M=M(s_{Min})$}

The second-type dipole test has already been applied on the SCP supernovae
of paper VI, where the adopted value of the Hubble Magnitude $M_B$ is that
minimizing the standard deviation of the unweighted least square fitting. In
this case we derive the value of the Hubble depth $D$, to be introduced in
eq. (14) for each sample analysed, as follows: 
\begin{equation}
\left[ \gamma ,z,m,M=M(s_{Min})\right] \Rightarrow D=\frac{10^{0.2\left[
m-M(s_{Min})\right] -5}}{1+z}\Rightarrow \left[ Y\rightarrow (-\cos \gamma
)\right] \Rightarrow a^{*}
\end{equation}

Here the test regards all 4 pilot samples XVI-XVII-XVIII-XIX of Table 0; in
particular the fitting according to (19) has been carried out both on the 4
primary samples of Table 1a (XVI$_1$ - XVII$_1$ - XVIII$_1$ - XIX$_1$: cf.
Table 1b), and on a further 18 encapsulated $z$ bins (XVI$_{2\rightarrow 10}$
- XVIII$_{2\rightarrow 10}$: cf. Tables 3-4).

\subsection{Solution}

All the sample features and results of the above $1^{st}$ and $2^{nd}$
dipole test appear in the following Tables 1ab-2-3-4.

Specifically, the main features of the 4 primary samples at $\langle
z\rangle =1.0$ are listed in Table 1a, with the following data: Sample
ordinal number; number N of supernovae of the sample; sample $z$ bin; mean $%
\langle z\rangle $ of the $z$ bin; unweighted mathematical mean $\langle
m_B^{\max }\rangle $ of the sample SNe magnitudes; mean $\langle \cos \gamma
\rangle $ of the sample SNe $\cos \gamma $.

\vspace{.1in}

\textbf{Table 1a}

\vspace{.1in}

\begin{tabular}{|c|c|c|c|c|c|}
\hline
Sample & N & $z$ bin & $\langle z\rangle $ & $\langle m_B^{\max }\rangle $ & 
$\langle \cos \gamma \rangle $ \\ \hline
XVI$_1$ & 48 & $0.830\leq z\leq 1.390$ & 1.001 & 24.84 & +0.29 \\ \hline
XVII$_1$ & 64 & $0.811\leq z\leq 1.400$ & 0.992 & 24.81 & +0.27 \\ \hline
XVIII$_1$ & 58 & $0.812\leq z\leq 1.400$ & 0.995 & 24.82 & +0.24 \\ \hline
XIX$_1$ & 62 & $0.812\leq z\leq 1.400$ & 0.996 & 24.80 & +0.27 \\ \hline
\end{tabular}
\ 

\vspace{.4in}

The dipole test on the above 4 primary samples of Table 1a was made possible
thanks to data and references taken from the SCP papers.

The fitting results are shown in Table 1b, where two $1^{st}$ type primary
sample tests and four $2^{nd}$ type primary sample tests are presented
numerically. In particular the $8$ columns of Table 1b present the following
data: Test identification name (TID); sample ordinal number; number N of the
sample supernovae; the fitting standard deviation $s$ in H.u. that results
either from the Hubble Magnitude values provided by the function $M(D)$
according to the paper IX parameters or as the minimum value of the standard
deviation, $s_{Min}$, corresponding to the listed Hubble Magnitude $%
M(s_{Min})$; the function $M(D)$ or the $M$ value which minimizes the
standard deviation $s$ in the dipole least square fitting; average Hubble
depth $\langle D\rangle $ of the sample SNe whose individual $D$ come from
the previous $M$ value; the resulting angular coefficient $a^{*}$ of eq.
(14) with its standard deviation; the correlated maximum $cz$ range of the
fitted sample, obtained by $c\left| \Delta z\right| ^{\max }=2a^{*}D$ with $%
D=4283$ $Mpc$ as the adopted Hubble depth at the central redshift $z_0=1.000$%
.

\newpage\ 

\textbf{Table 1b}

\vspace{.1in}

\begin{tabular}{|cccccccc|}
\hline
\multicolumn{1}{|c|}{TID} & \multicolumn{1}{c|}{Sample} & 
\multicolumn{1}{c|}{N} & \multicolumn{1}{c|}{$s$} & \multicolumn{1}{c|}{$M$}
& \multicolumn{1}{c|}{$\langle D\rangle $} & \multicolumn{1}{c|}{$a^{*}$} & $%
c\left| \Delta z\right| ^{\max }$ \\ \hline
\multicolumn{1}{|c|}{A1} & \multicolumn{1}{c|}{XVI$_1$} & 
\multicolumn{1}{c|}{48} & \multicolumn{1}{c|}{$10.374$} & 
\multicolumn{1}{c|}{$d_0+d_1D+d_2D^2$} & \multicolumn{1}{c|}{$4419$} & 
\multicolumn{1}{c|}{$6.0\pm 2.5$} & $\approx 51000$ \\ \hline
\multicolumn{1}{|c|}{B1} & \multicolumn{1}{c|}{XVI$_1$} & 
\multicolumn{1}{c|}{48} & \multicolumn{1}{c|}{$s_{Min}=7.2108$} & 
\multicolumn{1}{c|}{$M(s_{Min})=-19.89$} & \multicolumn{1}{c|}{$4477$} & 
\multicolumn{1}{c|}{$5.5\pm 1.7$} & $\approx 47000$ \\ \hline
\multicolumn{1}{|c|}{C1} & \multicolumn{1}{c|}{XVII$_1$} & 
\multicolumn{1}{c|}{64} & \multicolumn{1}{c|}{$18.654$} & 
\multicolumn{1}{c|}{$d_0+d_1D+d_2D^2$} & \multicolumn{1}{c|}{$4427$} & 
\multicolumn{1}{c|}{$4.8\pm 3.8$} & $\approx 41000$ \\ \hline
\multicolumn{1}{|c|}{D1} & \multicolumn{1}{c|}{XVII$_1$} & 
\multicolumn{1}{c|}{64} & \multicolumn{1}{c|}{$s_{Min}=12.779$} & 
\multicolumn{1}{c|}{$M(s_{Min})=-19.95$} & \multicolumn{1}{c|}{$4614$} & 
\multicolumn{1}{c|}{$5.9\pm 2.6$} & $\approx 51000$ \\ \hline
\multicolumn{1}{|c|}{E1} & \multicolumn{1}{c|}{XVIII$_1$} & 
\multicolumn{1}{c|}{58} & \multicolumn{1}{c|}{$s_{Min}=7.4327$} & 
\multicolumn{1}{c|}{$M(s_{Min})=-19.87$} & \multicolumn{1}{c|}{$4420$} & 
\multicolumn{1}{c|}{$5.1\pm 1.7$} & $\approx 44000$ \\ \hline
\multicolumn{1}{|c|}{F1} & \multicolumn{1}{c|}{XIX$_1$} & 
\multicolumn{1}{c|}{62} & \multicolumn{1}{c|}{$s_{Min}=7.7170$} & 
\multicolumn{1}{c|}{$M(s_{Min})=-19.89$} & \multicolumn{1}{c|}{$4432$} & 
\multicolumn{1}{c|}{$4.3\pm 1.7$} & $\approx 37000$ \\ \hline
\end{tabular}

\vspace{.4in}

Table 1b demonstrates the full success of this dipole test at $\langle
z\rangle =1.00$. Clearly the lower standard deviation $s$ indicate B1 and E1
as being the best test fittings. It is remarkable that the mathematical mean
of the 6 angular coefficients $a^{*}$ coming from the 6 primary tests of
Table 1b gives $\langle a^{*}\rangle =5.3\pm 0.3$ H.u. as a model
independent result, in full accordance with the value $a_{ECM}^{*}=5.46$
predicted by the expansion center model (cf. section 4). Consequently the
maximum $cz$ range of the resulting dipole nears a value of about $50000$ $%
km $ $s^{-1}$, as the last column of $c\left| \Delta z\right| ^{\max }$
certifies.

A further remark on Table 1b must be made about the scattering between the
resulting mean $\langle D\rangle $ and the value of the Hubble depth $D=$ $%
c/H_0$ at $z_0=1$. Such a scattering is clearly tied to the solution in the $%
5^{th}$ column, whose $M$ values in modulus result to be sensibly greater
(at minimum $\sim 0.05$ magnitudes) than both $M=\langle m_B^{\max }\rangle
-5\log (2c/H_0)-25\cong -19.82$ after introducing $\langle m_B^{\max
}\rangle $ $=24.84$ and $M=d_0+d_1D+d_2D^2=-19.83$ being $D=4283$ $Mpc$. Its
entity, as $\Delta D=\langle D\rangle -D$ with $\frac{\Delta D}D\approx 0.04$%
, may be due to some small systematic perturbation or distortion effect (cf.
section 3.1 of the parallel paper XVI)

The following Tables 2-3-4 present another 27 sample tests, divided into 3
groups of 9 encapsulated $z$ bins; each group comes from the primary sample
tests, A1, B1, E1 of Table 1b. Therefore the column contents mantain the
same headings as Table 1. However Table 2 lacks the $M$ column because of
its variable value as $M(D)$.

\newpage\ 

\textbf{Table 2}

\vspace{.1in}

\begin{tabular}{|c|c|c|c|c|c|c|c|}
\hline
TID & Sample & N & $z$ bin & $\langle z\rangle $ & $s$ & $a^{*}$ & $c\left|
\Delta z\right| ^{\max }$ \\ \hline
A2 & XVI$_2$ & 45 & $0.833\leq z\leq 1.37$ & 1.000 & 10.625 & $6.3\pm 2.6$ & 
$\approx 54000$ \\ \hline
A3 & XVI$_3$ & 42 & $0.84\leq z\leq 1.35$ & 0.999 & 10.703 & $7.1\pm 2.7$ & $%
\approx 61000$ \\ \hline
A4 & XVI$_4$ & 39 & $0.85\leq z\leq 1.31$ & 0.998 & 11.046 & $7.5\pm 2.9$ & $%
\approx 64000$ \\ \hline
A5 & XVI$_5$ & 36 & $0.86\leq z\leq 1.30$ & 0.998 & 11.443 & $7.1\pm 3.2$ & $%
\approx 61000$ \\ \hline
A6 & XVI$_6$ & 33 & $0.87\leq z\leq 1.27$ & 0.997 & 11.895 & $7.1\pm 3.3$ & $%
\approx 61000$ \\ \hline
A7 & XVI$_7$ & 30 & $0.88\leq z\leq 1.23$ & 0.996 & 11.791 & $6.3\pm 3.3$ & $%
\approx 54000$ \\ \hline
A8 & XVI$_8$ & 27 & $0.90\leq z\leq 1.20$ & 0.996 & 11.502 & $4.3\pm 3.5$ & $%
\approx 37000$ \\ \hline
A9 & XVI$_9$ & 24 & $0.92\leq z\leq 1.15$ & 0.995 & 11.474 & $2.0\pm 3.7$ & $%
\approx 17000$ \\ \hline
A10 & XVI$_{10}$ & 21 & $0.93\leq z\leq 1.13$ & 0.985 & 10.880 & $1.7\pm 3.8$
& $\approx 15000$ \\ \hline
\end{tabular}

\vspace{.6in}

\textbf{Table\ 3}

\vspace{.1in}

\begin{tabular}{|c|c|c|c|c|c|c|c|c|}
\hline
TID & Sample & N & $z$ bin & $\langle z\rangle $ & $s_{Min}$ & $M(s_{Min})$
& $a^{*}$ & $c\left| \Delta z\right| ^{\max }$ \\ \hline
B2 & XVI$_2$ & 45 & $0.833\leq z\leq 1.37$ & 1.000 & 7.385 & $-19.89$ & $%
5.7\pm 1.8$ & $\approx 49000$ \\ \hline
B3 & XVI$_3$ & 42 & $0.84\leq z\leq 1.35$ & 0.999 & 7.609 & $-19.88$ & $%
5.6\pm 1.9$ & $\approx 48000$ \\ \hline
B4 & XVI$_4$ & 39 & $0.85\leq z\leq 1.31$ & 0.998 & 7.756 & $-19.89$ & $%
6.5\pm 2.1$ & $\approx 56000$ \\ \hline
B5 & XVI$_5$ & 36 & $0.86\leq z\leq 1.30$ & 0.998 & 7.859 & $-19.89$ & $%
5.8\pm 2.2$ & $\approx 50000$ \\ \hline
B6 & XVI$_6$ & 33 & $0.87\leq z\leq 1.27$ & 0.997 & 8.179 & $-19.89$ & $%
5.7\pm 2.3$ & $\approx 49000$ \\ \hline
B7 & XVI$_7$ & 30 & $0.88\leq z\leq 1.23$ & 0.996 & 7.729 & $-19.91$ & $%
5.6\pm 2.2$ & $\approx 48000$ \\ \hline
B8 & XVI$_8$ & 27 & $0.90\leq z\leq 1.20$ & 0.996 & 7.395 & $-19.92$ & $%
4.2\pm 2.2$ & $\approx 36000$ \\ \hline
B9 & XVI$_9$ & 24 & $0.92\leq z\leq 1.15$ & 0.995 & 7.428 & $-19.93$ & $%
3.4\pm 2.4$ & $\approx 29000$ \\ \hline
B10 & XVI$_{10}$ & 21 & $0.93\leq z\leq 1.13$ & 0.985 & 6.786 & $-19.93$ & $%
4.0\pm 2.4$ & $\approx 34000$ \\ \hline
\end{tabular}

\newpage\ 

\textbf{Table 4}

\vspace{.1in}

\begin{tabular}{|c|c|c|c|c|c|c|c|c|}
\hline
TID & Sample & N & $z$ bin & $\langle z\rangle $ & $s_{Min}$ & $M(s_{Min})$
& $a^{*}$ & $c\left| \Delta z\right| ^{\max }$ \\ \hline
E2 & XVIII$_2$ & 54 & $0.817\leq z\leq 1.39$ & 0.998 & 6.787 & $-19.88$ & $%
4.6\pm 1.6$ & $\approx 39000$ \\ \hline
E3 & XVIII$_3$ & 50 & $0.823\leq z\leq 1.38$ & 1.001 & 6.478 & $-19.90$ & $%
5.9\pm 1.6$ & $\approx 51000$ \\ \hline
E4 & XVIII$_4$ & 46 & $0.833\leq z\leq 1.33$ & 0.993 & 6.616 & $-19.89$ & $%
6.0\pm 1.7$ & $\approx 51000$ \\ \hline
E5 & XVIII$_5$ & 42 & $0.845\leq z\leq 1.33$ & 1.007 & 6.689 & $-19.90$ & $%
7.1\pm 1.8$ & $\approx 61000$ \\ \hline
E6 & XVIII$_6$ & 38 & $0.860\leq z\leq 1.30$ & 1.000 & 6.574 & $-19.91$ & $%
6.8\pm 1.9$ & $\approx 58000$ \\ \hline
E7 & XVIII$_7$ & 34 & $0.873\leq z\leq 1.27$ & 1.003 & 6.740 & $-19.93$ & $%
7.3\pm 2.0$ & $\approx 63000$ \\ \hline
E8 & XVIII$_8$ & 30 & $0.885\leq z\leq 1.20$ & 0.995 & 5.646 & $-19.95$ & $%
5.7\pm 1.8$ & $\approx 49000$ \\ \hline
E9 & XVIII$_9$ & 26 & $0.920\leq z\leq 1.15$ & 0.998 & 5.770 & $-19.95$ & $%
4.8\pm 2.0$ & $\approx 41000$ \\ \hline
E10 & XVIII$_{10}$ & 22 & $0.930\leq z\leq 1.12$ & 0.983 & 5.078 & $-19.95$
& $4.9\pm 1.9$ & $\approx 42000$ \\ \hline
\end{tabular}

\vspace{.4in}

Again, the mathematical mean of the angular coefficient $a^{*}$ of these 27
further dipole tests is $\langle a^{*}\rangle =5.52\pm 0.30$ H.u., while all
33 $a^{*}$ values listed in Tables 1b-2-3-4 give the mean 
\begin{equation}
\langle a^{*}\rangle =5.47\pm 0.25\text{ }km\text{ }s^{-1}Mpc^{-1}
\end{equation}
as the final result of this model independent dipole test.

Indeed the resulting value $5.47$ of $\langle a*\rangle $ coincides
perfectly with the value $a_{ECM}^{*}=5.46$ furnished by the expansion
center model at the central redshift $z_0=1$ (cf. the next section 4.1).
Consequently, the new Hubble D law of eq. (7) gives a corresponding mean for
the maximum $cz$ range of the \textbf{SNe Ia }$\mathbf{cz}$\textbf{\ dipole }%
at $D=4283$ $Mpc$, that is 
\begin{equation}
\langle c\left| \Delta z\right| ^{\max }\rangle =2D\langle a^{*}\rangle
=46900\pm 2200\text{ }km\text{ }s^{-1}
\end{equation}

\newpage\ 

\subsection{Tabled data set of 2 SCP SNe Ia samples at $\langle z\rangle
=1.00$}

Tables 5a-5b-5c list a basic data set of this crucial dipole test of the
expansion center Universe; this base set, which concerns two primary samples
of Table 1a, XVI$_1$ and XVIII$_1$ - of 48 $SCPU$\ SNe Ia and 58 $SCPU2$\
SNe Ia respectively - , lists all the data needed to check ten $1^{st}$ type
dipole tests and twenty $2^{nd}$ type dipole tests. We must point out that
the last column of Tables 5a and 5b list the $Y$ values of the Test A1
supernovae, as $Y_{A1}=\frac{cz}D-H_0$ where $cz$ and $D$ appear in Table
3abcdefghi of the previous ECM paper IX, while Table 5c refers only to a
further 12 SNe Ia data, those necessary to complete the $2^{nd}$ type Test E1%
$\rightarrow $E10. These listed values of $Y_{A1}$ are also useful for
checking the other 9 dipole tests, corresponding to 9 encapsulated $z$ bins
and called A2-A3-A4-A5-A6-A7-A8-A9-A10 in Table 2. Of course the $Y$ values
of the B and E series do not appear in Table 5 because their values change
according to the variable $M(s_{Min})$ solution of each corresponding $%
2^{nd} $ type dipole test. Specifically, the columns of Table 5a and Table
5b report in order the following data: Supernova name according to the
reference number of column 4$^{th}$; right ascension $\alpha $ and
declination $\delta $ as given in the reference in column 4; footnote
reference number of the work in which the previous tabled values of right
ascension (R.A.) and declination (Decl.) appear; redshift $z$ of supernova
or host galaxy as listed in the SCP papers, but rounded off to the third
decimal place as the CMB reference likely affects the value for about 0.001
on average; supernova magnitudes $m_{SCPU}$ and $m_{SCPU2}$ as $m_B^{\max }$
values listed in the SCP Union compilation ($SCPU:$\ Kowalski et al. 2008)
and in SCP Union2 ($SCPU2:$\ Amanullah et al. 2010); $-\cos \gamma $ value
of the supernova according to eq. (16) computed using the supernova
astronomical coordinates $(\alpha ,\delta )$ listed on the Internet (ref. 1:
Harvard-IAU 2003) or in the reference papers cited in the footnote to Table
5c; $Y_{A1}$ value as above explained.

The dipole diagrams referring to the listed data in Table 5abc, those of the
Tests A1-B1-E1 of Table 1b, are reported in Figures 1-2-3, respectively.
These 3 primary sample tests are graphically presented by as many dipole
plots of $Y=\frac{cz}D-H_0$ against the ($-\cos \gamma $) value of each
corresponding supernova. In the cartesian plane $(x,y)$ of Figures 1-2-3 the
resulting fitting equations, as $y=f(x)$, are included together with the
value of the coefficient of determination $\mathbf{R}^2$. Note that the mean
of the 3 $a^{*}$, as $\langle a^{*}\rangle =5.5\pm 0.3$ H.u., coincides with
the mean (20).

\newpage\ 

\textbf{Table 5a}

\vspace{.1in}

\begin{tabular}{|c|c|c|c|c|c|c|c|c|}
\hline
Name & R.A. & Decl. & ref. & $z_{SCP}$ & $m_{SCPU}$ & $m_{SCPU2}$ & $-\cos
\gamma $ & $Y_{A1}$ \\ \hline
1997ck & 16 53.0 & +35 04 & 1 & 0.970 & 24.72 & 24.69 & $-0.0482$ & $+1.7$
\\ \hline
1997ap & 13 47.2 & +02 24 & 1 & 0.830 & 24.34 & 24.31 & $-0.2912$ & $+1.2$
\\ \hline
1999fm & 02 30.6 & +01 10 & 1 & 0.950 & 24.30 & 24.22 & $+0.1006$ & $+21.7$
\\ \hline
1999fk & 02 28.9 & +01 16 & 1 & 1.057 & 24.77 & 24.74 & $+0.1061$ & $+10.4$
\\ \hline
2002aa & 07 48.8 & +10 18 & 1 & 0.946 & 24.60 & ..... & $-0.9007$ & $+4.5$
\\ \hline
2002x & 08 48.5 & +44 16 & 1 & 0.859 & 24.73 & 24.68 & $-0.9684$ & $-12.1$
\\ \hline
2002w & 08 47.9 & +44 14 & 1 & 1.031 & 24.47 & ..... & $-0.9684$ & $+24.3$
\\ \hline
2001kd & 07 50.5 & +10 21 & 1 & 0.936 & 24.96 & 24.82 & $-0.9029$ & $-12.7$
\\ \hline
2001jm & 04 39.2 & -01 33 & 1 & 0.978 & 24.50 & 24.44 & $-0.3511$ & $+14.3$
\\ \hline
2001jh & 02 29.0 & +00 21 & 1 & 0.885 & 24.31 & 24.17 & $+0.1137$ & $+10.8$
\\ \hline
2001hu & 07 50.6 & +09 58 & 1 & 0.882 & 24.91 & 24.71 & $-0.9007$ & $-16.5$
\\ \hline
2001hs & 04 39.4 & -01 33 & 1 & 0.833 & 24.26 & 24.22 & $-0.3503$ & $+5.5$
\\ \hline
2001fs & 04 39.5 & -01 28 & 1 & 0.874 & 25.12 & 25.07 & $-0.3514$ & $-24.1$
\\ \hline
1997ek & 04 56.2 & -03 41 & 1 & 0.860 & 24.48 & 24.47 & $-0.3875$ & $-1.3$
\\ \hline
04Eag & 12:37:20.75 & +62:13:41.50 & 2 & 1.020 & 24.97 & 24.93 & $-0.6777$ & 
$-4.0$ \\ \hline
04Gre & 03:32:21.49 & -27:46:58.30 & 2 & 1.140 & 24.73 & 24.75 & $+0.1253$ & 
$+24.5$ \\ \hline
04Man & 12:36:34.81 & +62:15:49.06 & 2 & 0.854 & 24.53 & 24.55 & $-0.6786$ & 
$-4.4$ \\ \hline
04Mcg & 03:32:10.02 & -27:49:49.98 & 2 & 1.370 & 25.73 & 25.64 & $+0.1263$ & 
$-2.4$ \\ \hline
04Omb & 03:32:25.34 & -27:45:03.01 & 2 & 0.975 & 24.88 & 24.88 & $+0.1248$ & 
$-5.1$ \\ \hline
04Pat & 12:38:09.00 & +62:18:47.24 & 2 & 0.970 & 25.02 & 24.99 & $-0.6762$ & 
$-11.5$ \\ \hline
04Sas & 12:36:54.11 & +62:08:22.76 & 2 & 1.390 & 25.82 & 26.00 & $-0.6786$ & 
$-4.4$ \\ \hline
05Fer & 12:36:25.10 & +62:15:23.84 & 2 & 1.020 & 24.83 & 24.80 & $-0.6789$ & 
$+2.6$ \\ \hline
05Gab & 12:36:13.83 & +62:12:07.56 & 2 & 1.120 & 25.07 & 25.09 & $-0.6794$ & 
$+2.0$ \\ \hline
05Lan & 12:36:56.72 & +62:12:53.33 & 2 & 1.230 & 26.02 & 26.05 & $-0.6783$ & 
$-22.4$ \\ \hline
\end{tabular}

\newpage\ 

\textbf{Table 5b}

\vspace{.1in}

\begin{tabular}{|c|c|c|c|c|c|c|c|c|}
\hline
Name & R.A. & Decl. & ref. & $z_{SCP}$ & $m_{SCPU}$ & $m_{SCPU2}$ & $-\cos
\gamma $ & $Y_{A1}$ \\ \hline
05Red & 12:37:01.70 & +62:12:23.98 & 2 & 1.190 & 25.76 & 25.63 & $-0.6782$ & 
$-18.0$ \\ \hline
05Spo & 12:37:06.53 & +62:15:11.70 & 2 & 0.839 & 24.20 & 24.15 & $-0.6779$ & 
$+9.5$ \\ \hline
05Str & 12:36:20.63 & +62:10:50.58 & 2 & 1.010 & 25.03 & 24.92 & $-0.6793$ & 
$-7.7$ \\ \hline
2002dd & 12 36.9 & +62 13 & 1 & 0.950 & 24.66 & 24.61 & $-0.6784$ & $+2.0$
\\ \hline
2002fw & 03 32.6 & -27 47 & 1 & 1.300 & 25.65 & 25.65 & $+0.1244$ & $-5.3$
\\ \hline
2002hp & 03 32.4 & -27 46 & 1 & 1.305 & 25.41 & 25.51 & $+0.1250$ & $+6.1$
\\ \hline
2002ki & 12 37.5 & +62 21 & 1 & 1.140 & 25.35 & 25.37 & $-0.6770$ & $-7.8$
\\ \hline
2003az & 12 37.3 & +62 19 & 1 & 1.265 & 25.68 & 25.73 & $-0.6774$ & $-9.5$
\\ \hline
2003dy & 12 37.2 & +62 11 & 1 & 1.340 & 25.77 & 25.70 & $-0.6780$ & $-6.6$
\\ \hline
2003eq & 12 37.8 & +62 14 & 1 & 0.840 & 24.35 & 24.34 & $-0.6770$ & $+2.1$
\\ \hline
04D4bk & 22:15:07.681 & -18:03:36.79 & 3 & 0.840 & 24.31 & 24.32 & $+0.9345$
& $+4.1$ \\ \hline
04D3nr & 14:22:38.526 & +52:11:15.06 & 3 & 0.960 & 24.54 & 24.56 & $-0.4827$
& $+9.5$ \\ \hline
04D3ki & 14:19:34.598 & +52:17:32.61 & 3 & 0.930 & 24.87 & 24.69 & $-0.4885$
& $-9.8$ \\ \hline
04D3cp & 14:20:23.954 & +52:49:15.45 & 3 & 0.830 & 24.24 & 24.11 & $-0.4884$
& $+6.1$ \\ \hline
04D4dw & 22:16:44.667 & -17:50:02.38 & 3 & 0.961 & 24.57 & 24.55 & $+0.9317$
& $+8.1$ \\ \hline
04D3lp & 14:19:50.911 & +52:30:11.88 & 3 & 0.983 & 24.93 & 24.97 & $-0.4886$
& $-6.4$ \\ \hline
03D4cy & 22:13:40.441 & -17:40:54.12 & 3 & 0.927 & 24.72 & 24.55 & $+0.9347$
& $-3.7$ \\ \hline
03D1ew & 02:24:14.079 & -04:39:56.93 & 3 & 0.868 & 24.37 & 24.31 & $+0.1748$
& $+5.1$ \\ \hline
04D3dd & 14:17:48.411 & +52:28:14.57 & 3 & 1.010 & 25.12 & 24.88 & $-0.4931$
& $-11.4$ \\ \hline
03D4di & 22:14:10.249 & -17:30:24.18 & 3 & 0.905 & 24.29 & 24.24 & $+0.9335$
& $+15.0$ \\ \hline
03D4cx & 22:14:33.754 & -17:35:15.35 & 3 & 0.949 & 24.50 & 24.47 & $+0.9333$
& $+10.1$ \\ \hline
04D3ml & 14:16:39.095 & +53:05:35.89 & 3 & 0.950 & 24.55 & 24.51 & $-0.4976$
& $+7.6$ \\ \hline
04D3gx & 14:20:13.666 & +52:16:58.33 & 3 & 0.910 & 24.71 & 24.67 & $-0.4870$
& $-5.3$ \\ \hline
03D1cm & 02:24:55.294 & -04:23:03.61 & 3 & 0.870 & 24.46 & 24.54 & $+0.1699$
& $+1.0$ \\ \hline
\end{tabular}

\newpage\ 

\textbf{Table 5c}

\vspace{.1in}

\begin{tabular}{|c|c|c|c|c|c|c|}
\hline
Name & R.A. & Decl. & ref. & $z_{SCP}$ & $m_{SCPU2}$ & $-\cos \gamma $ \\ 
\hline
2001jf & 02 28.1 & +00 27 & 1 & 0.815 & 24.83 & $+0.1162$ \\ \hline
2001hy & 08 49.8 & +44 15 & 1 & 0.812 & 24.86 & $-0.9686$ \\ \hline
04D3lu & 14:21:08.009 & +52:58:29.74 & 1 & 0.822 & 24.34 & $-0.4872$ \\ 
\hline
04D3nc & 14:16:18.224 & +52:16:26.09 & 3 & 0.817 & 24.24 & $-0.4959$ \\ 
\hline
03D4cn & 22:16:34.600 & -17:16:13.55 & 3 & 0.818 & 24.63 & $+0.9297$ \\ 
\hline
1999fj & 02 28.4 & +00 39 & 1 & 0.816 & 24.17 & $+0.1134$ \\ \hline
2002fx & 03 32.1 & -27 45 & 1 & 1.400 & 25.65 & $+0.1258$ \\ \hline
2003aj & 03 32.7 & -27 55 & 1 & 1.307 & 26.97 & $+0.1253$ \\ \hline
2003XX & 12:37:29.00 & +62:11:27.8 & 4 & 0.935 & 24.48 & $-0.6776$ \\ \hline
2001cw & 15 23.1 & +29 40 & 1 & 0.953 & 24.71 & $-0.1718$ \\ \hline
2001gn & 14 02.0 & +05 05 & 1 & 1.124 & 25.37 & $-0.2603$ \\ \hline
2001hb & 13 57.2 & +4 20 & 1 & 1.030 & 24.79 & $-0.2715$ \\ \hline
\end{tabular}

\vspace{.2in}

R.A.\&Decl. references in Table 5abc:

$^1$Harvard-IAU, http://cfa-www.harvard.edu/iau/lists/Supernovae.html

$^2$Riess, A.G. et al. (2007) as referenced by Kowalski, M. et al. (2008)

$^3$Astier, P. et al. (2006) as referenced by Kowalski, M. et al. (2008)

$^4$Riess, A.G. et al. (2004) as referenced by Kowalski, M. et al. (2008)

\newpage

\section{The expansion center model reconfirmed}

The solution of the expansion center model (ECM) within the very nearby
Universe, using the 83 individual galaxies listed by Sandage \& Tammann in
their paper V (S\&T 1975) - at $z<0.02$ with $\langle z\rangle \equiv
z_0=0.0066$ or $D(z_0)=28.3$ $Mpc$ and single redshifts $z$ corrected only
for the Sun's motion in the Local Group through the standard vector of $300$ 
$km$ $s^{-1}$ towards $l=90^0;$ $b=0^0$ -, gives the values 
\begin{equation}
H_0=70\pm 3\text{ }km\text{ }s^{-1}Mpc^{-1}\hspace{.2in}a_0\cong 12.7\text{ }%
km\text{ }s^{-1}Mpc^{-1}\hspace{.2in}R_0=260\pm 22\text{ }Mpc
\end{equation}
as resulting from the Hubble ratio eq. (3) of the ECM paper II (Lorenzi
1999b), here rewritten as the formulation for the wedge-shape of the new
Hubble law, that is 
\begin{equation}
cz=\left[ H_0-a_{ECM}^{*}(D)\cdot \cos \gamma \right] \cdot D
\end{equation}
with 
\begin{equation}
cz=\dot r\hspace{.2in}a_{ECM}^{*}=a_0(1-x)^{\frac 13}/(1+x)\hspace{.2in}D=%
\frac{cx}{3H_0}\left( \frac{1+x}{1-x}\right) \hspace{.2in}x=\frac{3H_0r}c%
\hspace{.2in}a_0=\frac{3H_0^2R_0}c
\end{equation}
and 
\begin{equation}
\cos \gamma \equiv 0\Rightarrow z_0=\frac x3\left( \frac{1+x}{1-x}\right)
\end{equation}

\subsection{A few ECM values at $z_0=1$}

The direct extension and application of the previous formulae to the Deep
Universe with its central redshift $z_0=1$ leads to the approximate ECM
values, as follows: 
\begin{equation}
z_0=1\Rightarrow x^2+4x-3=0\Rightarrow x=0.6457513\Rightarrow r\cong 922%
\text{ }Mpc
\end{equation}
\begin{equation}
z_0=1\Rightarrow D=\frac c{H_0}\Rightarrow D\cong 4283\text{ }Mpc
\end{equation}
\begin{equation}
z_0=1\Rightarrow a_{ECM}^{*}(x)=a_{ECM}^{*}(D)\cong 5.46\text{ }km\text{ }%
s^{-1}Mpc^{-1}
\end{equation}
\begin{equation}
z_0=1\Rightarrow c\left| \Delta z\right| ^{\max }=2a_{ECM}^{*}\cdot D\cong
46770\text{ }km\text{ }s^{-1}
\end{equation}

In addition, recalling the formulae (cf. papers I-III) 
\begin{equation}
R=R_0(1-x)^{\frac 13}\hspace{.5in}t=t_0(1-x)\hspace{.5in}t_0=\frac 1{3H_0}
\end{equation}
we obtain the Galaxy distance $R$ from the expansion center at the epoch $t$
of $z_0=1$ : 
\begin{equation}
z_0=1\Rightarrow R=184\pm 16\text{ }Mpc\hspace{.1in}\text{at the epoch}%
\hspace{.1in}t\approx 1.7\times 10^9years
\end{equation}

\subsection{The ECM decelerating expansion}

In\ conclusion the whole of the results, collected without and within the
ECM in about a quarter of a century, give strong observational evidence for
an expansion center Universe, radially decelerated towards the huge void
center $VC$ ( $\alpha _{VC}\approx 9^h;\delta _{VC}\approx +30^0$: Bahcall
\& Soneira 1982), in full accordance with what is described within the 1999
ECM paper II. That scientific evidence coming from historic and recent
observational data sets shows the physical consistency of \textbf{a
decelerating expansion dipole}, detectable at any Hubble depth $D$, both in
the nearby and deep Universe, \textbf{with a resulting angular coefficient }$%
a^{*}$\textbf{or }$a_0$ of the linear fitting of the Hubble ratio $cz/D$
plotted versus $-\cos \gamma $ or $-X$, respectively.

The angular coefficient $a_0$, which was called \textbf{Galaxy radial
deceleration coefficient} (cf. paper II), is represented below through the
multiple formulation in Hubble units as follows 
\begin{equation}
a_0=K_0R_0=R_0\left( \frac{\delta H}{\delta r}\right) _{r=0}=\frac{3H_0^2R_0}%
c=\frac{-3c}2\left( \frac{\delta ^2R_{MW}}{\delta r^2}\right) _{r=0}km\text{ 
}s^{-1}Mpc^{-1}
\end{equation}

Here the Galaxy radial deceleration towards the expansion center $VC$, in $%
c.g.s.$ units, results to be 
\begin{equation}
\left( \frac{\delta ^2R_{MW}}{\delta t^2}\right) _{t=t_0}=-2H_0^2R_0\approx
-8.2\times 10^{-9}cm\text{ }s^{-2}
\end{equation}
corresponding to the value of the cosmic matter density $\rho _0$ at our
epoch $t_0$, given by the paper VII formula (21), that is 
\begin{equation}
\rho _0=\frac{3H_0^2n}{8\pi G_0}=2.3_{-0.5}^{+0.7}\times 10^{-28}g\text{ }%
cm^{-3}
\end{equation}
where $n=V_R/V_{ECM}=24.8_{-3.9}^{+4.7}$, while the Galaxy radial and
transversal velocities, after fixing the cosmic rotation $\dot \vartheta
_0=y_0H_0$ with the resulting $y_0=3.2_{-0.3}^{+0.4}$, take the values 
\begin{equation}
\dot R_0=H_0R_0\approx 1.8\times 10^9cm\text{ }s^{-1}\hspace{1.0in}R_0\dot
\vartheta _0\approx 6\times 10^9cm\text{ }s^{-1}
\end{equation}
respectively (cf. formulae and numerical values in the ECM paper VII).

\subsection{Measuring the deceleration parameter $q_0$}

The above sections 4.1 - 4.2, following the previous crucial dipole test
that has been made possible thanks to the Supernova Cosmology Project
(Perlmutter et. al. 1999 - Knop et al. 2003 - Kowalski et al. 2008 -
Amanullah et al. 2010), undertake another confirmation of the expansion
center model (ECM), specifically remarking the strong physical \textbf{%
evidence for a high cosmic deceleration }towards the expansion center, even
of the very nearby Universe, where the calculation of the decelerating
expansion dipole coefficient $a^{*}\cong a_0$, the controversial value of $%
H_0$ and the Galaxy radial run $R_0$ follow easily from historic distance
measures, such as those made over three decades from the $1950s$ by Hubble's
fellow scientist, Prof. \textbf{Allan Sandage} (Wikipedia 2011).

On the other hand, the very different context of relativistic observational
cosmology includes the problem of measuring the \textbf{deceleration
parameter }$q_0$, whose computed values of the past century seemed to span a
positive range, according to the analysis of the Hubble diagram of rich
clusters (e.g.: Sandage 1972 : $q_0\cong +1.0\pm 0.5$ - Rowan-Robinson 1996
: $q_0\cong +1.6\pm 0.4$). But in the first decade of the century the
measurement and confirmation of a resulting negative value of the
deceleration parameter ($\langle q_0\rangle \approx -1:$ John 2004), based
on the analysis of SCP high-$z$ SNe data within relativistic cosmology, led
to the discovery of the accelerating expansion through the observation of
distant supernovae.

However, it is possible to show that even the relativistic Universe of the
supernovae Ia is undergoing a highly decelerating expansion, that means
finding \textbf{a high positive value of }$q_0$, in accordance with the
expansion center model. These are the contents of the ECM paper XII,
presented at the meeting EWASS 2012 and here attached to the main contents
of paper X, as section 5 of the new paper XV.

\section{Evidence for a high deceleration of the relativistic Universe}

A new calculation of the relativistic deceleration parameter $q_0$ is the
topic of the present section, following the APPENDIX\ - November 2011 -
''Introduction to the Hubble Magnitude and a new relativistic $q_0$'' to the
ECM paper X (SAIt2011 in Palermo).

\subsection{Formulation of the SNe Ia absolute magnitude}

Regarding the supernovae of the SCP Union compilation (Kowalski et al.
2008), let us proceed through a few statements which refer to a $z$-bin
normal point of the redshift $z$, the corresponding intrinsic luminosity $L$
and the apparent magnitude $m$, within a classical model-independent
cosmology. While the $z$-bin normal redshift $\langle z\rangle $, like the
normal apparent magnitude $\langle m\rangle $, is an observed quantity $O$,
and $z_0$, like $m_0=m(z_0)$, is a calculated central value $C$ (cf. papers
IX-XVI), here $O-C=0$ is assumed to hold for a \textbf{normalized-central
supernova Ia;} that means the following: 
\begin{equation}
z=\langle z\rangle \equiv z_0
\end{equation}
\begin{equation}
L\equiv L(z_0)=\alpha L_0\hspace{.2in}with\hspace{.2in}\alpha =\alpha
(z_0)\geq 1
\end{equation}
\begin{equation}
z_0\rightarrow 0\Rightarrow L=L_0
\end{equation}
\begin{equation}
m=\langle m\rangle \equiv m_0=-2.5\log \frac L{4\pi d_L^2}+const.=-2.5\log
(\alpha L_0)+5\log D_L+30+C
\end{equation}
with $C=const.+2.5\log (4\pi )$ and $d_L=D_L\times 10^6$ $\equiv d_L(z_0)$
parsecs.

The absolute magnitude has by convention the formulation 
\begin{equation}
M_\alpha =-2.5\log \frac L{4\pi \cdot 10^2}+const.=-2.5\log (\alpha L_0)+5+C
\end{equation}

So we have 
\begin{equation}
M_\alpha =m-5\log D_L-25=M_0-2.5\log \alpha
\end{equation}
being 
\begin{equation}
M_0=-2.5\log L_0+5+C
\end{equation}
Hence the difference between the SNe Ia \textbf{intrinsic absolute magnitude 
}$M_\alpha $ and the hypothetical absolute magnitude $M_0$ of the same
source located at $10$ parsecs at our epoch $t_0$ can be defined as the $%
M_\alpha $\textbf{\ spread}, by the formula 
\begin{equation}
M_\alpha -M_0=-2.5\log \alpha
\end{equation}

\subsection{Relativistic Magnitude $M_R(z_0)$}

Once taken into account eq. (41) as the classical fomulation of the absolute
magnitude in cosmology, the first problem is to give a correct formula to
the \textbf{luminosity distance} $D_L$. In relativistic cosmology we have 
\begin{equation}
D_L=D_0\cdot (1+z_0)
\end{equation}
where $D_0=D_{pr}(t_0)$\textbf{\ represents the proper distance} in
Megaparsecs at the present epoch $t=t_0$ \textbf{of a supernova Ia,} with
redshift $z=\langle z\rangle \equiv z_0$, according to the eqs.
(36)(37)(38)(39). As the relativistic formula of $D_0$ (cf. Coles \& Lucchin
1995) is the following 
\begin{equation}
D_0=\frac{cz_0}{H_0}\left[ 1-\frac{z_0}2(1+q_0)+...\right]
\end{equation}
we obtain 
\begin{equation}
D_L=\frac{cz_0}{H_0}\left[ 1+\frac{z_0}2(1-q_0)+...\right]
\end{equation}
where $c$ is the velocity of light in $km/s$, $H_0=H(t_0)$ the Hubble
constant at $t=t_0$ in $km/s/Mpc$ and $q_0$ a dimensionless deceleration
parameter, that can be written without the relativistic scale factor or
expansion parameter, as follows 
\begin{equation}
q_0=-\frac{\ddot D_{pr}(t_0)\cdot D_{pr}(t_0)}{\dot D_{pr}^2(t_0)}=-\frac{%
\ddot D_0D_0}{\dot D_0^2}
\end{equation}
after recalling the \textbf{relativistic Hubble law} 
\begin{equation}
\dot D_{pr}(t_0)=H_0D_0
\end{equation}
and expanding the proper distance $D_{pr}(t)$ at the epoch $t$ in a
power-series: 
\begin{equation}
D_{pr}(t)=D_0\cdot \left[ 1+H_0(t-t_0)-\frac 12q_0H_0^2(t-t_0)^2+...\right]
\end{equation}

The introduction of $D_L$ eq. (46) into $M_\alpha (z_0)$ eq. (41) leads to
the formulation of the \textbf{relativistic absolute magnitude} $M_R(z_0)$,
that is 
\begin{equation}
M_\alpha (z_0)\equiv M_R(z_0)=m_0-25+5\log H_0-5\log cz_0-1.0857\cdot
(1-q_0)\cdot z_0+...
\end{equation}

It is important to remark that, owing to the assumed constancy of the SNe Ia
mean intrinsic luminosity, at least for $z_0\ll 1$ or $\alpha (z_0)=1$, the
previous \textbf{relativistic magnitude }$M_R(z_0)$\textbf{\ of a supernova
Ia}, with $z=\langle z\rangle \equiv $ $z_0$ and $m=\langle m\rangle \equiv
m_0$ assumed, must be practically coinciding with the relativistic absolute
magnitude $M_0$ of a hypothetical supernova Ia with a redshift $%
z_0\rightarrow 0$. Therefore with $z_0\ll 1$, for instance at $z_0=0.001$
and within the observational limits, we can write 
\begin{equation}
M_0=m_0-25+5\log H_0-5\log cz_0-1.0857\cdot (1-q_0)\cdot z_0
\end{equation}

At the same time, if we assume the constancy of $M_R(z_0)$ as $M_0$ at any $%
z_0$ according to eq. (50), the observation of many distant SNe Ia with
different $z_0$ allows us to compute the deceleration parameter $q_0$. A
negative value of $q_0$ implies the accelerating expansion of the
relativistic Universe, as discovered by the Nobel scientists Saul
Perlmutter, Brian Schmidt and Adam Riess (2011).

\subsection{Hubble Magnitude $M(z_0)$ and its total spread}

The relativistic Hubble law (48) can be replaced by the alternative law 
\begin{equation}
cz_0=H_0D
\end{equation}
where now $D=\frac{cz_0}{H_0}$ \textbf{is an apparent distance}, which
differs from the proper distance $D_0=D_{pr}(t_0)$ according to eq. (45). $D$%
\ is\textbf{\ }the \textbf{Hubble depth} of a central point $z_0$ (cf.
papers IX-XVI). Consequently, as in eq. (44), here a different luminosity
distance $D_C$ (that is the ECM cosmological distance) based on the Hubble
depth $D$ can be formulated as follows 
\begin{equation}
D_C=D\cdot (1+z_0)=\frac{cz_0}{H_0}(1+z_0)
\end{equation}

At this point, the introduction of $D_L=$ $D_C$ in eq. (41) leads to the
mathematical definition of the \textbf{Hubble Magnitude }$M(z_0)$\textbf{\
of a supernova Ia}, with $z=\langle z\rangle \equiv $ $z_0$ and $m=\langle
m\rangle \equiv m_0$ assumed: 
\begin{equation}
M(z_0)=m_0-5\log \left[ D\cdot (1+z_0)\right] -25
\end{equation}
or 
\begin{equation}
M(z_0)=m_0-25+5\log H_0-5\log cz_0-5\log (1+z_0)
\end{equation}

One can remark that the previous eq. (53) of $D_C$ results to be practically
coinciding with the relativistic $D_L$ eq. (46) at the second order (cf.
Attilio Ferrari 2011), that is at $z_0\ll 1$, after introducing the obtained
value $q_0\approx -1$ ($\langle q_0\rangle \approx -0.77:$ John 2004) from
the SNe Ia observational cosmology; as a consequence, also the eq. (54) or
(55) of the Hubble Magnitude $M(z_0)$ results to coincide with the
relativistic formulation of the absolute magnitude $M_\alpha (z_0)\equiv
M_R(z_0)$ at $z_0\ll 1$ (cf. eq. (41) and (50)). At the same time such a $%
q_0\approx -1$ is the result of the assumed constancy of $M_R(z_0)$ as $M_0$
at any $z_0$, that means $\alpha =1$ assumed in eq. (41). So the
relativistic cosmology implies that the Hubble Magnitude $M(z_0)$ at $z_0\ll
1$ coincides with the relativistic absolute magnitude $M_0$ of eq. (51) at $%
z_0\rightarrow 0$.

Hence the relativistic result can be summarized through the following two
statements: 
\begin{equation}
M_R(z_0)\equiv M_0\text{ at any }z_0\Rightarrow q_0\approx -1
\end{equation}
\begin{equation}
q_0=-1\hspace{.1in}at\hspace{.1in}z_0\ll 1\Rightarrow
M(z_0)=M_R(z_0)=M_0\Rightarrow M(z_0)-M_0=0
\end{equation}

The previous $M(z_0)-M_0$ is the total $M$\ spread, that is the deviation of
the Hubble Magnitude $M(z_0)$,\ of a supernova Ia with redshift $z=\langle
z\rangle \equiv z_0$, from the $M_0$ value corresponding to the same $z_0$.
The below eq. (58) gives the\textbf{\ relativistic expression of that total }%
$M(z_0)$\textbf{\ spread:} 
\begin{equation}
M(z_0)-M_0=-5\log (1+z_0)+1.0857\cdot (1-q_0)\cdot z_0+...
\end{equation}

It is remarkable that the previous eq. (58) was obtained from the difference
between eq. (55) and the $M_0$ eq. (51), with the consequent elimination
both of $H_0$ and $m_0$.

\subsection{Computation of the deceleration parameter $q_0$}

From eq. (58), only for $z_0\ll 1$, the formula for \textbf{the deceleration
parameter }$q_0$ becomes 
\begin{equation}
q_0\cong +1-4.605\times z_0^{-1}\times \log (1+z_0)-0.921\times
z_0^{-1}\times \left[ M(z_0)-M_0\right]
\end{equation}

Eq. (59), that is eq. A19 of the APPENDIX\ - November 2011 - to the ECM
paper X, appears to be the solution to an apparent paradox in relativistic
cosmology. In fact, when $M(z_0)=M_0$ at $z=z_0\ll 1$ is assumed, eq. (59)
still gives the negative value $q_0\approx -1$ in accordance with the Deep
Universe analysis carried out by the Nobel scientists Saul Perlmutter, Brian
Schmidt and Adam Riess (2011), while the total $M$ spreads which can be
inferred from the SCP Union data give $q_0$ a positive value, according to
the preliminary results obtained in the cited Appendix of paper X. On this
occasion, the ''absolute magnitude analysis of the SCP Union supernovae'' of
the parallel paper XVI makes it possible to calculate and extrapolate a few
cubic fittings of the SNe Hubble Magnitude $M$ versus $z=\langle z\rangle
\equiv z_0$ and the correlated spreads $\left[ M-M_0\right] $ at $z_0\ll 1$,
with $\langle M\rangle \equiv M(z_0)\equiv M$ assumed, according to the
equation 
\begin{equation}
M=A_0+A_1z_0+A_2z_0^2+A_3z_0^3
\end{equation}
where $A_0\equiv M_0$. Let us remark that the extrapolated trends of the
normal or central Hubble Magnitude $M$\ of the supernovae Ia at low central
redshifts $z_0\equiv \langle z\rangle \ll 1$\ have a sharp negative
variation, which clearly contrasts with the almost constant trend due to a
relativistic $q_0\approx -1$. In particular the fitting solutions,
(41)(55)(59)(70) of paper XVI and graphically represented by the fit lines
of the Appendix Figures 14-22-24-29 of the same parallel paper, give four
total $M$ spreads, whose numerical values with the resulting $q_0$ from eq.
(59) are collected below in Table 6, Table 7, Table 8 and Table 9,
respectively. In particular the values of Table 9 refer to the final $M$
solution (70) of paper XVI, with $M_0=-17.9$ .

\vspace{.1in}

\hspace{.5in}\textbf{Table 6\hspace{0.78in}Table 7}\hspace{0.78in}\textbf{%
Table 8}\hspace{0.75in}\textbf{Table 9}

\vspace{.05in}

\begin{tabular}{|ll|ll|ll|lll|}
\hline
\multicolumn{1}{|l||}{$z_0$} & $M-M_0$ & \multicolumn{1}{|l||}{$q_0$} & $%
M-M_0$ & \multicolumn{1}{|l||}{$q_0$} & $M-M_0$ & \multicolumn{1}{|l||}{$q_0$%
} & \multicolumn{1}{l|}{$M-M_0$} & $q_0$ \\ \hline
\multicolumn{1}{|l||}{$0.001$} & $-0.00411$ & \multicolumn{1}{|l||}{$+2.79$}
& $-0.00335$ & \multicolumn{1}{|l||}{$+2.09$} & $-0.00319$ & 
\multicolumn{1}{|l||}{$+1.93$} & \multicolumn{1}{l|}{$-0.00426$} & $+2.92$
\\ \hline
\multicolumn{1}{|l||}{$0.01$} & $-0.04085$ & \multicolumn{1}{|l||}{$+2.77$}
& $-0.03325$ & \multicolumn{1}{|l||}{$+2.07$} & $-0.03161$ & 
\multicolumn{1}{|l||}{$+1.92$} & \multicolumn{1}{l|}{$-0.04229$} & $+2.91$
\\ \hline
\multicolumn{1}{|l||}{$0.1$} & $-0.3807$ & \multicolumn{1}{|l||}{$+2.60$} & $%
-0.3096$ & \multicolumn{1}{|l||}{$+1.95$} & $-0.2929$ & 
\multicolumn{1}{|l||}{$+1.79$} & \multicolumn{1}{l|}{$-0.3946$} & $+2.73$ \\ 
\hline
\end{tabular}

\vspace{.3in}

Clearly, the main result here presented is $q_0\gtrsim +2$, which shows 
\textbf{relativistic cosmology expressing a positive and high value of the
deceleration parameter }$q_0$.

\subsection{The relativistic $q_0$ from 249 high-$z$ SNe Ia according to ECM}

More rigorously, within the ECM context, the cubic fitting (60) for the
calculation of the total $M$ spread to introduce into eq. (59) should be
applied exclusively to individual points of central Hubble Magnitude, that
is to say only $M(z_0)$ values based on central $m_0$ and $z_0$ in eq. (55)
which refer to individual supernovae with $\cos \gamma \approx 0$ (cf. eqs.
18 in paper IX). Alternatively, on the grounds of the dipole analysis
carried out on 249 high-$z$ SCP Union supernovae according to the expansion
center model, one can fit 249 Hubble Magnitudes $M_z$ plotted against the
ECM Hubble depth $D_z$ (cf. Table 3abcdefghi of papers XI-XVI), as shown in
the diagram of Figure 4, which replicates Figure 6 of paper XVI. In this
case, after applying 
\begin{equation}
D_z=cz/H_X=cz_0/H_0=D
\end{equation}
with $H_0=70$ $km/s/Mpc$ assumed (cf. paper I and II), the cubic line (60)
fitting the 249 individual points $M_z$ becomes 
\begin{equation}
M_z(z_0)=A_0+A_1z_0+A_2z_0^2+A_3z_0^3=d_0+d_1D+d_2D^2+d_3D^3=M_z(D)
\end{equation}
being 
\begin{equation}
A_0=d_0\hspace{.3in}A_1=d_1c/H_0\hspace{.3in}A_2=d_2c^2/H_0^2\hspace{.3in}%
A_3=d_3c^3/H_0^3
\end{equation}

The solution, from the automatic fitting based on the Hubble depth $D=D_z$ ,
gives 
\begin{equation}
d_0=-18.15\hspace{.3in}d_1=-9.05E-04\hspace{.3in}d_2=+1.69E-07\hspace{.3in}%
d_3=-1.15E-11
\end{equation}
where $\mathbf{R}^2=0.382$ is the value of the coefficient of determination.

The above cubic solution, reported as $y=f(x)$ in the Figure 4 area,
together with the value of $\mathbf{R}^2$, gives the total $M$ spread a more
reliable extrapolated value at $z_0=0.001$ or $D=4.28$ $Mpc$, the following

\[
\left[ M(z_0)-M_0\right] =\left[ M_z(z_0)-A_0\right] =-0.00387\hspace{.2in}at%
\hspace{.05in}z_0=0.001 
\]
, whose introduction in the formula (59) leads to a high value of the
relativistic $q_0$, that is 
\begin{equation}
q_0=+2.57
\end{equation}

\newpage\ 

\subsection{Concordance test on Galaxy radial deceleration}

The section ''Verso una nuova cosmogonia della concordanza'' of the ECM
paper VIII, ''Steps towards the expansion center cosmology'', represents the
first step of a search for some points of contact between the ECM and
relativistic cosmology. Here a new concordance test may take into account
the calculation of the Galaxy radial deceleration in paper VII, based on the
main ECM motion equations. In particular eq. (41) of paper I, that is 
\begin{equation}
\ddot R_{MW_{t=t_0}}=-2H_{(s^{-1})}^2(t_0)\cdot R_{(cm)}(t_0)
\end{equation}
becomes perfectly equivalent to the relativistic 
\begin{equation}
\ddot D_0=-q_0H_0^2D_0
\end{equation}
that follows from the previous eqs. (47)(48), when $R_0=D_0$ is considered
to be the proper distance at $t_0$\ of the expansion center from the Galaxy.
In this case the expansion center model gives the relativistic deceleration
parameter $q_0$ the value of $+2$.

Now the calculation in $c.g.s.$ units of the relativistic deceleration $%
\ddot D_0$ of eq. (67), with $q_0=+2$ or $q_0=+3$ applied respectively,
after adopting the ECM values $H_0=69.8\pm 2.8$ $km$ $s^{-1}Mpc^{-1}$ based
on data by Sandage \& Tammann (1975) and $D_0\approx 260$ $Mpc$ as the
Galaxy distance from the huge void center (Bahcall \& Soneira 1982) (cf.
papers I-II-VII and author 1991), leads to the values listed below, in Table
10.

\vspace{.1in}

\textbf{Table 10}

\vspace{.07in}

\begin{tabular}{|c|c|c|c|}
\hline
$H_0$ & $D_0$ & $\ddot D_0(q_0=+2)$ & $\ddot D_0(q_0=+3)$ \\ \hline
$(2.263\pm 0.091)\times 10^{-18}s^{-1}$ & $\approx 8.0\times 10^{26}cm$ & $%
-0.8\times 10^{-8}cm/s^2$ & $-1.2\times 10^{-8}cm/s^2$ \\ \hline
\end{tabular}

\vspace{.4in}

The results reported in Table 10 can be summarized through a single Galaxy
radial deceleration $\ddot R_0$ $\equiv \ddot D_0$, according to the
following order of magnitude: 
\begin{equation}
\ddot R_0\approx -10^{-8}cm/s^2
\end{equation}

\newpage\ 

\section{Conclusion}

In conclusion the above multiple dipole test on SCP SNe at $\langle z\rangle
=1.0$ represents a last crucial proof of the expansion center Universe.
Moreover the expansion center model and the involved Galaxy deceleration are
here fully reconfirmed at Hubble depths of the Deep Universe. That called
for a further ECM dipole analysis on all 249 High-$z$ SCP Union supernovae,
those here referred to as pilot sample XVI in Table 0 and listed in paper IX
as usable supernovae with $z>0.2$ of the SCP Union compilation (Kowalski et
al.2008). A dipole and absolute magnitude analysis of these $SCPU$ SNe Ia
within the expansion center model has been carried out in the parallel paper
XVI, with important consequences for observational and relativistic
cosmology. In particular a new positive value of the relativistic parameter $%
q_0$ comes out from the extrapolated behaviour of supernovae Ia at redshift $%
z=\langle z\rangle \equiv z_0=0.001$. In other words even the expansion of
the relativistic Universe is shown to be decelerating, through a data
analysis referring to the nearby instead of the deep Universe. At the same
time such an analysis follows from extrapolated fittings of high-$z$ SNe Ia $%
M$ normal or central points, which have the advantage of being negligibly
affected by the perturbation due to cosmic rotation (cf. parallel paper XVI).

In the author's view, the previous results represent the overcoming of the
relativistic paradox of the accelerating cosmic expansion and its so-called
dark energy. Consequently here relativistic cosmology says Einstein was
right (Einstein-de Sitter 1932) when he rejected the cosmological constant,
whose introduction was the greatest blunder of his life, as he remarked much
later, according to George Gamow (1970).

\newpage\ 

\section{Acknowledgements}

This work has been made possible thanks to the SCP Union compilation. The
author would like to thank all the members of the SCP team, in particular
for making the SNe data available on line in ''arXiv:0804.4142v1 $\left[ 
\text{astro-ph}\right] $ 25 Apr 2008'' , ''arXiv:1004.1711v1 $\left[ \text{%
astro-ph.CO}\right] $ 10 Apr 2010'' and
''http://www-supernova.lbl.gov/Union/''.

On this occasion the author wishes to express his gratitude to the President
of ''Think Tank Security (TTS)'' , Prof. Nazzareno Mandolesi, jointly to the
astronomy professors Fabio Finelli e Paolo Natoli, for the useful and
stimulating meeting of the 23$^{th}$ June 2011 at the TTS centre in Rome
about the expansion center model and the CMB dipole anisotropy.

Special acknowledgements are reserved for the Local Organizing Committee of
the successful European Week of Astronomy and Space Science (EWASS 2012),
organized by the Italian Astronomical Society (SAIt) on behalf of the
European Astronomical Society (EAS) and held in Rome, Vatican City, at the
Pontifical Lateran University, during the week of July 1-6, 2012.

\newpage\ 

\section{References}

\vspace{.2in}

\hspace{.2in}Amanullah, R. et al. 2010, arXiv:1004.1711v1 $\left[ \text{%
astro-ph.CO}\right] $ 10 Apr 2010$\rightarrow $ApJ 716, 712

Bahcall, N.A. and Soneira, R.M. 1982, ApJ 262, 419 (B\&S)

Coles,P. \& Lucchin,F. 1995,''COSMOLOGY''-The Relativistic Universe-John
Wiley \& Sons

Einstein, A. \& de Sitter, W. 1932, Proc. N.A.S. 18, 213

EWASS 2012, http://www.ifsi-roma.inaf.it/ewass2012/

Ferrari, A. 2011, ''Fondamenti di astrofisica''- Parte VI : Cosmologia -
Springer Verlag Italia

Gamow, G. 1970, My World Line: An Informal Autobiography - Viking Press

Harvard-IAU 2003, http://cfa-www.harvard.edu/iau/lists/Supernovae.html

John, M. V. 2004, ApJ 614, 1

Knop, R.A. et al. 2003, ApJ 598, 102 (K03)

Kowalski, M. et al. 2008, arXiv:0804.4142v1 $\left[ \text{astro-ph}\right] $
25 Apr 2008$\rightarrow $ApJ 686, 749

Lorenzi, L. 1989, 1991, Contributi N. 0,1, Centro Studi Astronomia -
Mondov\`\i , Italy

\hspace{.5in}1993, in 1995 MemSAIt, 66, 1, 249

\hspace{.5in}1994, in 1996 Astro. Lett. \& Comm., 33, 143 (eds.-SISSA ref.
155/94/A)

\hspace{.5in}1999a, astro-ph/9906290 17 Jun 1999,

\hspace{.5in}in 2000 MemSAIt, 71, 1163 (paper I : reprinted in 2003,
MemSAIt, 74)

\hspace{.5in}1999b, astro-ph/9906292 17 Jun 1999,

\hspace{.5in}in 2000 MemSAIt, 71, 1183 (paper II : reprinted in 2003,
MemSAIt, 74)

\hspace{.5in}2002, in 2003 MemSAIt, 74, 480 (paper III-partial version),

\hspace{.5in}http://sait.oat.ts.astro.it/MSAIt740203/PDF/poster/39\_lorenzil%
\_01\_long.pdf

\hspace{.5in}(paper III-integral version)

\hspace{.5in}2003a, MemSAIt Suppl. 3, 277 (paper IV)

\hspace{.5in}2003b, MemSAIt Suppl. 3,

\hspace{.5in}http://sait.oat.ts.astro.it/MSAIS/3/POST/Lorenzi\_poster.pdf
(paper V)

\hspace{.5in}2004, MemSAIt Suppl. 5, 347 (paper VI: partial and integral
version)

\hspace{.5in}%
2008,http://terri1.oa-teramo.inaf.it/sait08/slides/I/ecmcm9b.pdf (paper VII)

\hspace{.5in}2009,
http://astro.df.unipi.it/sait09/presentazioni/AulaMagna/08AM/lorenzi.pdf

\hspace{.5in}(paper VIII)

\newpage\ 

\hspace{.5in}2010, arXiv:1006.2112v3 $\left[ \text{physics.gen-ph}\right] $
17 Jun 2010 (paper IX)

\hspace{.5in}2011a,
http://www.astropa.unipa.it/SAIT2011/Proceedings/Lorenzi1.pdf (paper X)

\hspace{.5in}2011b,
http://www.astropa.unipa.it/SAIT2011/Proceedings/Lorenzi2.pdf (paper XI)

\hspace{.5in}2012a, poster paper presented at EWASS 2012 (paper XII)

\hspace{.5in}2012b, parallel poster paper presented at EWASS 2012 (paper
XIII)

\hspace{.5in}2012e (paper XVI: parallel paper)

Perlmutter, S., et al. 1999, ApJ 517, 565 (P99)

Perlmutter, S., Schmidt, B.P., Riess, A.G. 2011, \textbf{Nobel Prize in
Physics 2011}

Rowan-Robinson, M. 1996, ''Cosmology'' Claredon Press - Oxford

Sandage, A. 1972, ApJ 178, 1

Sandage, A., Tammann G.A. 1975a, ApJ 196, 313 (S\&T: Paper V)

\end{document}